\documentstyle[11pt,ysc,twoside,epsf]{article}

\def\edcomment#1{\iffalse\marginpar{\raggedright\sl#1\/}\else\relax\fi}

\reversemarginpar

\begin{document}

\title{ Interstellar C2 Molecule as Seen in HST/STIS Data }

\author{Marcin Dyrka }

\affil{ Jan D\l ugosz Academy, Institute of Physics,
Cz\c{e}stochowa, Al. Armi Krajowej 13/15, 42-200 Cz\c{e}stochowa,
Poland.}

\author{Bogdan Wszo\l ek}

\affil{ Jan D\l ugosz Academy, Institute of Physics,
Cz\c{e}stochowa, Al. Armi Krajowej 13/15, 42-200 Cz\c{e}stochowa,
Poland, Jagiellonian University Astronomical Observatory, ul. Orla
171, 30-244 Krak\'{o}w, Poland \\ E-mail:bogdan{@}ajd.czest.pl }

\author{Micha\ l  Pawlikowski}

\affil{ Jagiellonian University Astronomical Observatory, ul. Orla 171, 30-244 Krak\'{o}w, Poland}

\begin{abstract}

Carbon chains are sometimes considered as possible carriers of some
diffuse interstellar bands. Spectroscopic observations in UV band
carried by spectrometer STIS fed with HST, give us the possibility
to detect many interstellar molecules. We focused our attention on
C2 molecule and we detected it in spectra of three reddened stars
(HD27778, HD147933, HD207198). Interstellar molecule C2 was detected
as a set of absorption lines around 2313 angstroms.

\end{abstract}

\section{Introduction}

Spectroscopic observations of the early spectral type stars in
visible light give us a rich astrophysical information about the
nature of Diffuse Interstellar Bands (DIBs). DIBs were first
detected about 85 year's ago by Heger (1922). From the beginning
they are the subject of intensive examination and analysis. Up
today, these absorption structures of interstellar origin are not
identified. We still do not know what the carriers of DIBs are
(Herbig-1995, Wszo\l ek and God\l owski-2003).

Some authors (e.g. Fulara 1993) claim that DIBs may originate due to
interaction of light with interstellar molecules named carbon chains
(like C2, C3, C4,...). To verify the hypothesis about carbon chains
as DIBs' carriers one needs to examine mutual correlation between
intensities of DIBs' and lines given by these molecules.

First detection of short carbon chains in interstellar clouds
announced in literature, mobilized us to check whether diffuse
clouds producing DIBs contain C2 - which probably is the most
abundant molecule among interstellar carbon chains.

\section{ Observational Data }

We did make use from spectroscopic UV data given by Space Telescope
Imaging Spectrograph (STIS) at HST. The access to HST data archive
is possible by visiting homepage: http://archive.stsci.edu/. We have
got UV spectra for our sample of reddened stars. This sample
contained 12 stars (HD): 22591, 23180, 24398, 24534, 27778, 34078,
147933, 192639, 198478, 206267, 207198 and 210839.

Archive spectra are accessible as binary FITS files. Spectra were
achieved with use of echelle technique therefore each FITS file is
divided into many orders.

Furthermore in some cases observations were made many times.

After decoding STIS data, using software package IRAF, there became
clear that retrieved spectra meet our criteria (quality, wave length
band, number of observing repetitions) only in three cases, namely
for stars: (HD) 27778, 147933 and 207198.

\section{ Data Analysis }

Three software packages were used for data analysis and presentation:

IRAF - helped us to achieve ASCII files (lambda, intensity) from original FITS files.

REWIA v. 1.4 - sofware package written by Jerzy Borkowski (Nicolaus
Copernicus Astronomical Center, Polish Academy of Sciences,
Toru\'{n}, Poland) was used for normalizing spectra and for dividing
them by continuum. Averaging procedure was carried out also with the
help of this program.

Finally, ORIGIN package was used for graphic presentation of spectra.

We focused our attention on D-X (2313 angstroms) band of C2
molecule. This band contains a set of numerous and well separated
rotational lines, named as Mulliken System. We have used
high-resolution STIS spectra with R=110000. Each star from our 3 -
element sample was observed 8 times. To maximize S/N ratio we
averaged observations and finally we achieved satisfying result.

We detected D-X band of C2 in spectra of all our stars. Figure 1 is
to show how good is detection of this band for our target stars.
Nineteen rotational lines of considered band is well visible. In the
case of HD207198 the Doppler splitting of lines is seen. That means
that we have to deal with more then one cloud on the line of sight
and that these clouds have different radial velocities.

\begin{figure}
\plotone{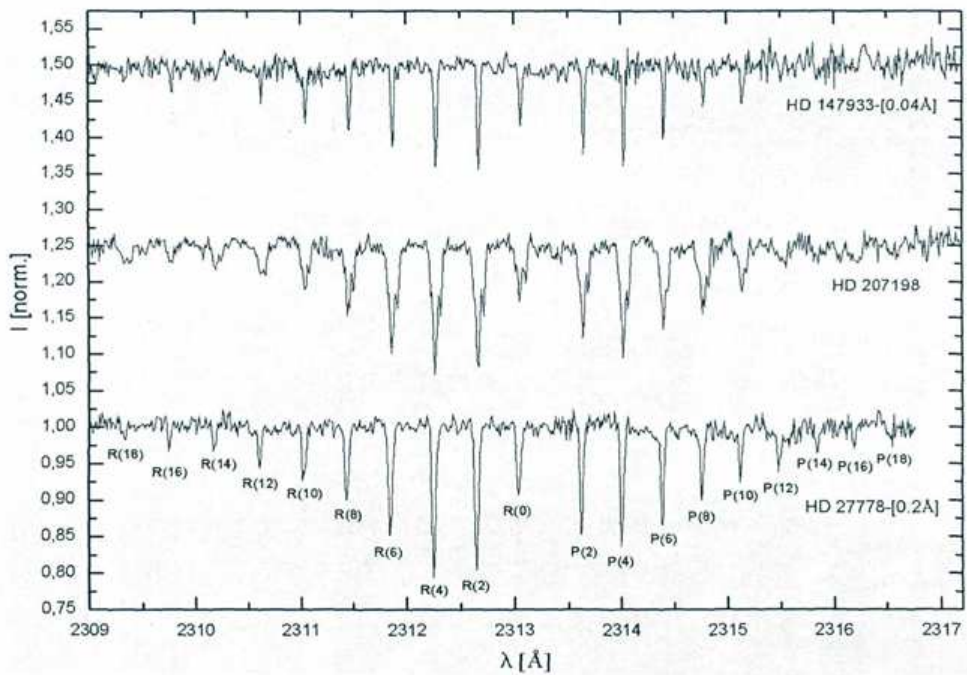}

\caption{STIS/HST spectra of reddened stars HD147933, HD207198 and
HD27778 in the region of C2 D-X (2313 angstroms) band. In
description of spectrograms there are given in brackets values of
small shifts of two spectra, which were done to draw all
spectrograms in common frame}
\end{figure}

\section{Conclusions}

The most important result of our analysis is that C2 molecules are
present in those interstellar clouds, which produce DIBs.
Furthermore 2313 angstroms band of C2 is easily detectable in STIS
data. Unfortunately STIS observations are very inhomogeneous
(different wavelength regions, different resolutions, different
gratings) and they make some difficulties when we want to acquire
numerous sample of spectra covering such wavelength region as we
wish. A sample of target stars with DIBs' detection counts about
100. From the other hand the only few stars of this sample has
satisfactory C2 detention. To solve the problem whether C2 may be
a crucial molecule as far as DIBs' carriers are concerned, one
needs much more observations of C2 lines.

\begin {references}

\reference Fulara J.: 1993, Nature 366, 433;

\reference Heger M.L.: 1922, Lick. Obs. Bull., 10, p.146;

\reference Herbig G.H.: 1995, ARA\&A, 33, p.19;

\reference Wszo\l ek B., God\l owski W.: 2003, MNRAS, 338, p.990;

\end {references}

\end{document}